\documentclass[aps,pra,groupedaddress,notitlepage]{revtex4-1}

\usepackage{amssymb}
\usepackage{amsmath}
\usepackage{amsfonts}
\usepackage{amsthm}
\usepackage{graphicx}
\usepackage{xcolor}
\usepackage{bm}
\usepackage{url}
\usepackage{hyperref}
\usepackage{cases}

\begin{document}

\title{Using analogues to investigate error growth and predictability in turbulence}
\author{Carlos Granero Belinchon$^{1,2}$\email{carlos.granero-belinchon@imt-atlantique.fr}, Mickael Bourgoin$^3$\email{mickael.bourgoin@ens-lyon.fr}, B\'ereng\`ere Dubrulle$^4$\email{berengere.dubrulle@cea.fr}}

\affiliation{
$^1$ Department of Mathematical and Electrical Engineering, IMT Atlantique, Lab-STICC, UMR CNRS 6285, 655 Av. du Technop\^ole, Plouzan\'e, 29280, Bretagne, France. \\
$^2$ Odyssey, Inria/IMT Atlantique, 263 Av. G\'en\'eral Leclerc, Rennes, 35042, Bretagne, France. \\
$^3$ CNRS, ENS de Lyon, LPENSL, UMR5672, 69342, Lyon cedex 07, France\\
$^4$ Universit\'e Paris-Saclay, CEA, CNRS, SPEC, 91191, Gif-sur-Yvette, France}

\date{\today}

\begin{abstract}
In this study, we revisit Lorenz's pioneering concept of analogues to investigate error growth and predictability in fully developed turbulence using experimental data.
By applying the method to a 1D turbulent velocity time series, we successfully identify the three error growth regimes intuited by Lorenz and later formalized by turbulence theory: the chaotic exponential regime, the algebraic regime of spontaneous stochasticity, and the saturation regime.
Our analysis reveals that the Lyapunov exponent and the exponent of the algebraic regime depend on the initial error. We interpret this dependence as a statistical bias induced by the intermittency of turbulence, where rare events, characterized by lower Lyapunov exponents and more singular velocity increments, are preferentially sampled when considering larger initial errors.
These findings not only validate Lorenz's qualitative intuition but also provide new quantitative insights into the predictability of turbulent flows, paving the way for potential applications in the prediction of extreme events.
\end{abstract}

\maketitle

\section{Introduction}
In the 1960's, Ed Lorenz set the basics of predictability in weather forecast in three milestone papers, introducing successively
the notion of chaos (sensitivity to initial conditions)~\cite{Lorenz1963}, spontaneous stochasticity (butterfly effect)~\cite{Lorenz1969} and analogues~\cite{Lorenz1969a}. These notions are rooted in the non-linearity of the primitive equations of fluid motion, which pilots error growth following two stages: (i) a {\sl chaotic} stage, based upon a linear mechanism, where the error growth is exponentially amplified; and (ii) a {\sl stochastic stage}, based upon a non-linear mechanism, where the error growth is algebraic. The first stage is a continuous mechanism, in the sense that the error is always proportional to the initial error. The second stage is discontinuous in the initial conditions, as it is independent of them.

After Lorenz's discovery of chaos, there were many attempts to determine the typical error-doubling time using numerical simulations of large-scale circulation models of the atmosphere~\cite{Smagorinsky1963,Mintz1964,Leith1965,Charney1966}, with an estimate of 5 days. However, as shown by Lorenz himself~\cite{Lorenz1969}, this doubling time is of the order of a few minutes for smaller structures like thunderstorms, shedding doubts on the validity of the concept of ``doubling-time'' without reference to the scales.
The very nature of the error growth, propagating from small to large scales in finite time, is rather suggestive of an inverse cascade of errors, terminating at the largest scale. This means that the large-scale doubling time is only an upper limit of error growth for structures of any size.

In 1969, large-scale circulation models had very limited resolution, setting a limit to the exploration of error growth, as the minimal error was necessarily bounded by the error at the resolution scale. This motivated an attempt by Lorenz to investigate the problem using observational data. He observed that pressure maps display recurring patterns with a degree of similarity, which he termed ``analogues''. He then attempted to study error growth between two different analogues as a function of the initial error between them. His idea was to prove, for example, that mediocre analogues (characterized by a rather large initial error) should be closer to the non-linear regime and thus exhibit a milder error growth than excellent analogues (with a small initial error). Due to limited observations, he could not verify this intuition in a systematic way. Nevertheless, this concept of analogues paved the way to several new lines of research, e.g. the prediction of extreme weather events~\cite{Yiou2014}, or the attribution of extreme events to climate change~\cite{Faranda2022}. On the theoretical side, Crisanti et al.~\cite{crisantiIntermittencyPredictabilityTurbulence1993, crisantiPredictabilityVelocityTemperature1993} later showed, within the multifractal formalism, that the intermittency of turbulence makes the Lyapunov exponent and the predictability time themselves statistically distributed quantities, whose scaling with the Reynolds number is affected by rare, highly dissipative events; this is an early hint of the kind of intermittency-induced bias that we investigate experimentally in the present paper.

In this paper, we come back to the original idea by Lorenz and use analogues to probe the issues of predictability in turbulence using experimental data. After a summary of theoretical results regarding error growth behavior and scaling, we explain why the analogue method is interesting in the context of experimental data and discuss its limitations. Then, we introduce the analogues method and we apply it to a 1D experimental turbulent velocity time series, showing the different error growth regimes intuited by Lorenz and formalized by turbulence theory, and extracting their characteristic quantities.

\section{Theoretical results on error growth}\label{sec:theo}

\subsection{General framework}

Error growth typically refers to a ``Cauchy picture'' of predictability, in which one compares the time-behavior of two solutions of the fluid equations $u_1(x,t)$ and $u_2(x,t)$ with initial conditions differing by a fixed initial amount $\delta_0$. Most of the time, the comparison is done using a simple $L_2$ norm, so that the error growth is monitored by:

\begin{equation}
\delta(t)\equiv \vert\vert u_1-u_2\vert\vert_{L_2}^2=\int d^3x \vert  u_1(x,t)-u_2(x,t)\vert^2,
\label{errorgrowthdef}
\end{equation}

\noindent with $\delta(t=0)=\delta_0$. In a turbulent flow, there are typically three stages for the time variation of $\delta$, as illustrated by~\cite{Thalabard2020}. At very short times, the error growth is exponential, characteristic of the chaotic regime:

\begin{equation}
\delta(t)\sim \delta_0\exp(\lambda t),
\label{chaotic}
\end{equation}

\noindent where $\lambda$ is the Lyapunov exponent. At very long times, the two solutions are fully decorrelated, so that the error growth saturates to:

\begin{equation}
\delta(t)\sim\vert\vert u_1\vert\vert_{L_2}^2+\vert\vert u_2\vert\vert_{L_2}^2=2u_{rms}^2,
\label{saturation}
\end{equation}

\noindent with $u_{rms}^2$ the flow root mean square (rms). In between these two regimes, there is an algebraic regime, where we have:

\begin{equation}
\delta(t)\sim \epsilon (t+t_b)^\alpha,
\label{SpSt}
\end{equation}

\noindent where $t_b$ is a typical time depending on the onset of the algebraic regime. Here, both the prefactor $\epsilon$ and the exponent $\alpha$ depend on the flow and, more precisely, on the specific observable considered: the velocity-based error used throughout this paper and the mean-square separation between fluid particles, for instance, are governed by the same turbulent cascade yet exhibit different exponents ($\alpha=1$ versus $\alpha=3$), as detailed in Section~\ref{sec:spont} below. Neither $\epsilon$ nor $\alpha$, however, depends on the initial error $\delta_0$. This regime is therefore discontinuous with respect to initial conditions, highlighting a breaking of determinism. It is therefore called the regime of spontaneous stochasticity.

The duration of the regimes and their features depend on the flow properties, with a number of interesting scaling laws.

\subsection{Chaotic stage scaling}

This regime is typically observed when $\delta(t)\ll 2u_{rms}^2$. Ruelle~\cite{Ruelle1979} predicted that the Lyapunov exponent $\lambda$ piloting the chaotic exponential growth of error should scale like the shortest time. In a velocity field with Holder exponent $h$, this results in $\lambda\sim Re^{(1-h)/(1+h)}$~\cite{Berera2018, Ho2020, Banerjee2026}. In incompressible 3D turbulence, $h=1/3$, so that one expects $\lambda\sim \sqrt{Re}$~\cite{Ruelle1979}. This is, however, seldom observed in numerical simulations of turbulence, where values of the exponent ranging from $0.53$ to $0.66$ have been reported~\cite{Berera2018, Ho2020, Boffetta1997, Ge2025}. One obvious explanation could be that turbulence is intermittent, which, in the multifractal picture~\cite{Frisch1985}, means that it is not uniquely characterized by $h=1/3$. However, taking into account the whole possible range of $h$ only results in a small negative correction with respect to Ruelle's scaling, $\lambda\sim Re^{0.493}$~\cite{crisantiIntermittencyPredictabilityTurbulence1993, crisantiPredictabilityVelocityTemperature1993}. More recently,~\cite{Banerjee2026} argued that the Lyapunov exponent scaling is monitored by the largest compressive eigenvalue of the strain $\lambda_3$, a claim supported by recent local investigation of the spatial distribution of the error~\cite{Ge2025}. Furthermore,~\cite{Banerjee2026} show that $\lambda_3$ fluctuates in time around a mean value which obeys the Ruelle scaling: $\left\langle \lambda_3 \right\rangle \sim\sqrt{Re} $. However, given that its standard deviation (std) obeys a stronger scaling $\sqrt{\left\langle \lambda_3^2 \right\rangle}\sim Re^{0.59}$, this results in a Lyapunov exponent $\lambda\sim -\left\langle \lambda_3 \right\rangle+\sqrt{\left\langle \lambda_3^2 \right\rangle}$ growing with a larger exponent than Ruelle's scaling. These predictions concern how $\lambda$ varies with the Reynolds number across different flows; in the present study, performed at a single, fixed Reynolds number, we instead probe a complementary source of variability, namely the quality $\delta_0$ of the analogue, and interpret it, following Crisanti et al.~\cite{crisantiIntermittencyPredictabilityTurbulence1993, crisantiPredictabilityVelocityTemperature1993}, as another manifestation of the same intermittency-driven scatter in the Lyapunov exponent.

\subsection{Spontaneous stochasticity}\label{sec:spont}

This regime is typically observed when $\delta(t)\sim u_{rms}^2$, corresponding to the beginning of a non-linear regime, which sets the onset time $t_b$. Physically, it originates from the non-smoothness of the turbulent velocity field: for a field with Holder exponent $h<1$, i.e. $\delta u(r)\sim r^h$, the flow map is not Lipschitz, so that two trajectories initially separated by an infinitesimal distance are no longer guaranteed to remain close. Assuming a Richardson-type closure, $dr/dt\sim \delta u(r)\sim r^h$, integrates to $r(t)\sim t^{1/(1-h)}$: the separation reaches a finite value in finite time, independently of how small the initial separation was: this is precisely the breaking of determinism introduced above.

The value of the exponent $\alpha$ then depends on which observable is built from this separation. For the velocity difference itself and in homogeneous turbulence without mean flow, $\delta u(t)\sim r(t)^h\sim t^{h/(1-h)}$, so that $\delta(t)=\delta u(t)^2\sim t^{2h/(1-h)}$, i.e. $\alpha=2h/(1-h)$~\cite{Dubrulle2026}. In 3D incompressible turbulence, the Kolmogorov value $h=1/3$ gives $\alpha=1$, with $\epsilon$ corresponding to the local energy dissipation~\cite{Thalabard2020}. For the particle separation itself, rather than the velocity difference, the same argument instead gives $r^2(t)\sim t^{2/(1-h)}$, i.e. an exponent of $3$ for $h=1/3$, the classical Richardson-Obukhov law for turbulent pair dispersion~\cite{Richardson1926}. The two cases illustrate that $\alpha$ is set by the observable considered, not solely by the flow.

\subsection{Towards experiments: analogues}

The above discussion shows that it would be interesting to check these results using experimental data. A key limitation is the inability to precisely control the initial conditions and subsequently reproduce the experiment under identical conditions with a prescribed perturbation. However, we can take advantage of ergodicity, meaning that the flow ``explores naturally'' the whole phase space over a sufficiently long time. In this framework, two fields differing by an amount $\delta_0$ (in $L_2$ norm) are expected to exhibit statistically indistinguishable dynamics over sufficiently short time intervals, up to deviations induced by the error growth governed by the underlying equation of motion. If this picture is true, we can then repeat what has been done numerically, using ``analogues'' as initial conditions. The main goal of the present paper is to test this idea.

\subsection{Limitations of the analogue approach}

However, this intuition is hindered by the curse of dimensionality. Indeed, as shown by~\cite{Cecconi2012}, for a given interval of time, the number of analogues within a given $\delta_0$ decreases exponentially with the number of degrees of freedom (and with the value of $\delta_0$). This rules out the use of 3D data sets, which are too limited in time extent to allow us to span a wide range of initial separations $\delta_0$. Therefore, we decided to use a simpler time-resolved data set, provided by the Modane experiment~\cite{Kahalerras1998}. This data set consists of the measurement of one component of the velocity field at a given location, $u(t)$. This simplicity therefore requires a subtle definition of analogues, which we describe now.

\section{Methods}
\subsection{Definition of an analogue}

Given a state $U$, belonging to a given phase space $\Omega$, we define its analogues up to a resolution $\delta_*$ as any state $V$ belonging to the same phase space and such that $d^2(U,V)\le \delta_*$, where $d$ is a distance in $\Omega$. This definition is valid for any phase space. In the sequel, we will work in a low-dimensional phase space that we will build by applying a Poincar\'e projection on the studied time series. The search for analogues then relies on two steps: (i) projection of the signal onto a finite-dimensional phase space $\Omega$ and (ii) for each observed state $U$, search for analogues $V$ at a given resolution $\delta_*$.

\subsection{Description of the data set}\label{sec:dataset}

The analyzed time series is an experimental turbulent velocity signal acquired in a wind tunnel using hot-wire anemometry~\cite{Kahalerras1998}. This method provides high temporal resolution, with response times of the order of tens of microseconds, enabling the capture of small-scale turbulent fluctuations. A passive grid installed upstream of the probe generates nearly homogeneous and isotropic turbulence.

Measurements were performed in the S1MA wind tunnel at ONERA Modane. The Taylor-scale Reynolds number of the flow is $R_{\lambda} \approx 2500$, which corresponds to a fully developed turbulence regime. The mean flow velocity is $\left\langle u\right\rangle \approx 20.5$ m/s. The sampling frequency is $f_s = \frac{1}{dt} = 25$ kHz, and the total duration is $1000$ s, yielding a time series $u(t)$ with $t\in\{t_1,\dots,t_N\}$, of approximately $N=2.5 \times 10^7$ samples. The integral and Kolmogorov time scales of the flow are respectively $T = 2350dt$ and $\tau_{K} = 5dt$. In the sequel, we construct the normalized velocity fluctuation by removing the mean and normalizing by the rms value:
$v(t)=(u- \bar{u} )/u_{rms}$.

\subsection{Poincar\'e projection}

Given the turbulent velocity time series $v(t)$ sampled homogeneously at time intervals $dt$, we use Poincar\'e projection, also called Takens embedding~\cite{Takens1981, Eckmann1985}, to build a $p$-dimensional phase space where the state at time $t$ is defined as:

\begin{equation}
    \vec{v}^{(p)}(t) = 
    \begin{pmatrix}
        v(t) \\
        v(t-\tau_*) \\
        \vdots \\
        v(t-(p-1)\tau_*)
    \end{pmatrix}.
\end{equation}

\begin{figure}[ht!]
\includegraphics[width=0.7\linewidth]{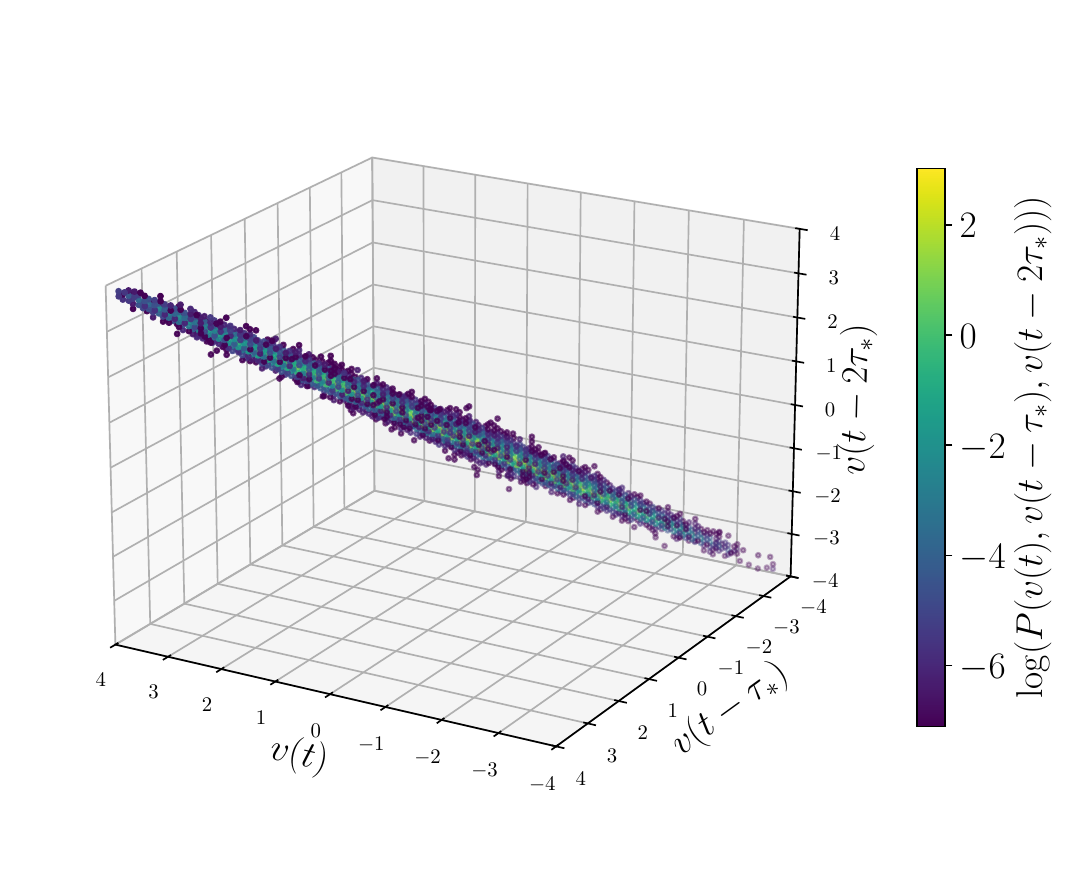}
 \centering
\caption{Phase space reconstructed using Poincar\'e projection on the Modane turbulent velocity time series with $p=3$ and $\tau_*=dt$.}\label{fig:space}
\end{figure}

This phase space construction depends on two parameters: the dimension of the phase space $p$ and the time sampling of the projection $\tau_*=m \, dt$ with $m \in \mathbb{N}$. In this work, we fix $\tau_*=dt$ and we study $p=3$, $p=10$ and $p=15$. The attractor of the three-dimensional phase space is shown in Figure~\ref{fig:space}.

\subsection{Analogues retrieval}

Given a state $\vec{v}^{(p)}(t)$ in the $p$-dimensional phase space reconstructed using Poincar\'e projection, and using the Euclidean distance $d$ in this phase space~\cite{Lguensat2017, Platzer2021}, the most obvious way to generate the ensemble of analogues at resolution $\delta_0$ is to construct the ensemble ${\cal A}(\vec{v}^{(p)}(t),\delta_0) = \left\lbrace \vec{v}^{(p)}(t'_i)\vert \delta_0 <d^2(\vec{v}^{(p)}(t),\vec{v}^{(p)}(t'_i))<2\delta_0  \right\rbrace$. If $\delta_0$ is very small, the cardinal of ${\cal A}(\vec{v}^{(p)}(t),\delta_0)$ is very small for any $\vec{v}^{(p)}(t)$, possibly $0$. As $\delta_0$ is increased, it is easier to find analogues, so the cardinal of ${\cal A}(\vec{v}^{(p)}(t),\delta_0)$ increases. The difficulty is, however, that the number of analogues depends on the state $\vec{v}^{(p)}(t)$, so that it is difficult to define controlled statistics.

To remedy this problem, we construct ensembles of analogues by fixing the cardinal of the ensemble. For this, we first define $\delta_0^{(k)}$ as the smallest value of $\delta_0$ such that the cardinal of ${\cal A}(\vec{v}^{(p)}(t),\delta_0)$ is equal to $k$, and note ${\cal C}(\vec{v}^{(p)}(t),k,0)={\cal A}(\vec{v}^{(p)}(t),\delta_0^{(k)})\equiv \left\lbrace \vec{v}^{(p)}(t'_i) \right\rbrace_{1\leq i \leq k}$. This defines the ensemble of analogues of $\vec{v}^{(p)}(t)$ as its $k$-nearest neighbors. With this definition, for each state $\vec{v}^{(p)}(t)$ and for fixed $k$, we associate a state-adaptive resolution $\delta_0^{(k)}$. Each analogue $\vec{v}^{(p)}(t'_i)$ has an associated successor $\vec{v}^{(p)}(t'_i+\tau)$ at time $\tau$. Therefore, we can define the ensemble of analogue successors at time $\tau$ as ${\cal C}(\vec{v}^{(p)}(t),k,\tau) = \left\lbrace \vec{v}^{(p)}(t'_i+\tau) \right\rbrace_{1\leq i \leq k}$.

In practice, we define a historical database in which to search for analogues, and an additional, independent realization used to sample the phase space, such that $t'_i < t$. We call these realizations {\sl database} and {\sl measure} respectively. For the $p$-dimensional phase space to be considered densely sampled, and so that the $k$ analogues are sufficiently close, we need a sufficiently long database for $v(t)$. In addition, the higher the dimension of the phase space, the more difficult it is to find analogues as $\delta_0\to 0$, and the larger the database has to be to define reasonable analogues for each state. 

The full experimental turbulent velocity signal presented in Section~\ref{sec:dataset} spans 1000 seconds, corresponding to 10638 integral scales. The first $419$ seconds, containing $N_D=5 \times 2^{21}$ samples, define the database used to search for analogues. This database contains approximately $4460$ integral scales. The measure realization is defined for times $t \in [838,922]$ seconds, corresponding to $N_M=2^{21}$ samples. Leaving a gap of 419 seconds between the database and the measure ensures that no correlations remain between them. Unless stated otherwise, we fix $k=50$ for $p=3$, and $k=200$ for $p=10$ and $p=15$.

\section{Error growth}

\subsection{Definition of error}

The previous construction provides a set of ``reference states'' (the $\vec{v}^{(p)}$ in the measure data set), with their successors, and their corresponding analogues (in the database) and analogue successors. From these, a natural definition of error would be the average over the analogues of the pairwise Euclidean distance between the reference state and its analogues, namely:

\begin{equation}\label{eq:errordefnatural}
\delta_0(\vec{v}^{(p)}) = \frac{1}{k}\sum_{i=1}^{k} \left(\vec{v}^{(p)}(t)-\vec{v}^{(p)}(t'_i) \right)^2,   
\end{equation}

\noindent and equivalently, the definition of the error at later times would be the average over the successors.

In practice, the interpretation of this definition of the error is ``biased'' towards the reference state, while this state is no more representative of the underlying physical state than any of its analogues, all of which are statistically equivalent. Motivated by this, we can define a smoother version of the error by computing the average of all the mean errors between each analogue and the rest of the members of the ensemble. This leads to:

\begin{equation}\label{eq:errordefana}
\delta_0(\vec{v}^{(p)}) = \frac{2}{k(k-1)}\sum_{i=1}^{k}\sum_{j>i} \left(\vec{v}^{(p)}(t'_i)-\vec{v}^{(p)}(t'_j) \right)^2,
\end{equation}

\noindent where the restriction $j>i$ ensures that we do not count the same distance twice.

Equivalently, we can define the error after a delay $\tau$ as the pairwise Euclidean distance between the $k$ analogue successors:

\begin{equation}\label{eq:errordefsuc}
\delta(\vec{v}^{(p)},\tau) = \frac{2}{k(k-1)}\sum_{i=1}^{k}\sum_{j>i} \left(\vec{v}^{(p)}(t'_i+\tau)-\vec{v}^{(p)}(t'_j+\tau) \right)^2.    
\end{equation}

At each time, this provides us with about $N_M\sim 2^{21}$ measurements, from which various statistical tests can be performed.

\subsection{Dependence on the initial error}

\begin{figure}[ht!]
\includegraphics[width=0.6\linewidth]{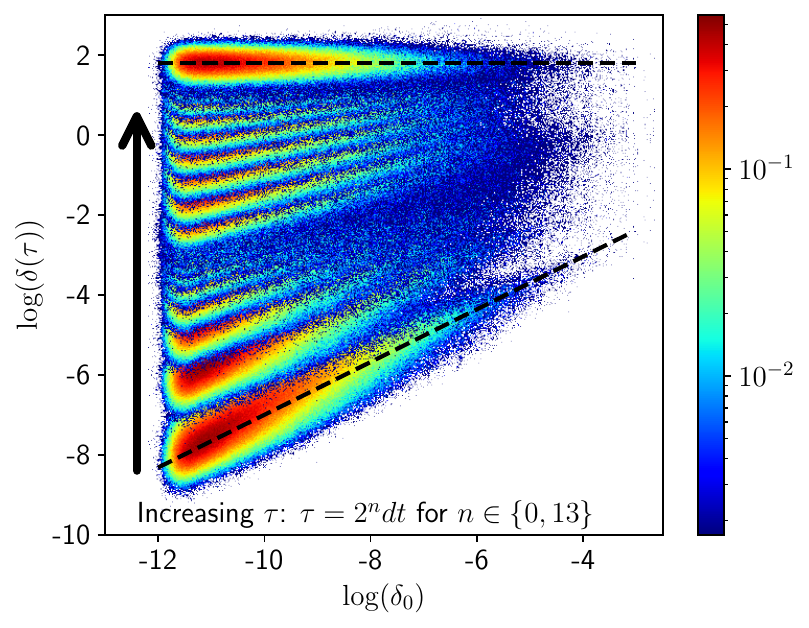}
 \centering
\caption{Logarithm of the error $\delta(\tau)$ as a function of the logarithm of the initial error $\delta_0$ for different values of $\tau$. A total of 14 histograms are shown corresponding respectively to $\tau=2^{n} dt$ with $n \in \left\lbrace 0,13 \right\rbrace$.}\label{fig:4}
\end{figure}

A first interesting result concerns the dependence of the error after a time $\tau$ as a function of the initial error $\delta_0$. This can be explored using the joint probability distribution $P(\delta_0,\delta(\tau))$ at different fixed time delays $\tau$. The results for 14 different values of $\tau=2^{n} dt$ with $n \in \left\lbrace 0,13 \right\rbrace$ are shown in Figure~\ref{fig:4}. A first striking result is that, at small $\tau$, the error is correlated with the initial error, while it becomes completely decorrelated from the initial error at the largest $\tau$. This is a qualitative illustration of the transition between the chaotic, deterministic regime (where error is proportional to the initial error) and the spontaneous stochastic regime, where the error no longer depends on the initial error. One can also observe on this graph that excellent analogues (with smaller initial errors) result in the strongest error growth with respect to mediocre analogues; a tendency already intuited by Lorenz in 1969. To check whether our framework bears more analogies with the classical error growth framework discussed in Section~\ref{sec:theo}, we now turn to more quantitative analysis.

\subsection{The error growth regimes}

Having established that the joint statistics of $\delta_0$ and $\delta(\tau)$ are consistent with the expected transition between the chaotic and spontaneous-stochastic regimes, we now examine how this transition unfolds in time, in order to identify and characterize the different error growth regimes predicted in Section~\ref{sec:theo}. The results shown in this section are obtained with $k=50$ and $k=2$ analogues and in a $3$-dimensional projection of the phase space.

\begin{figure}[ht!]
\includegraphics[width=\linewidth]{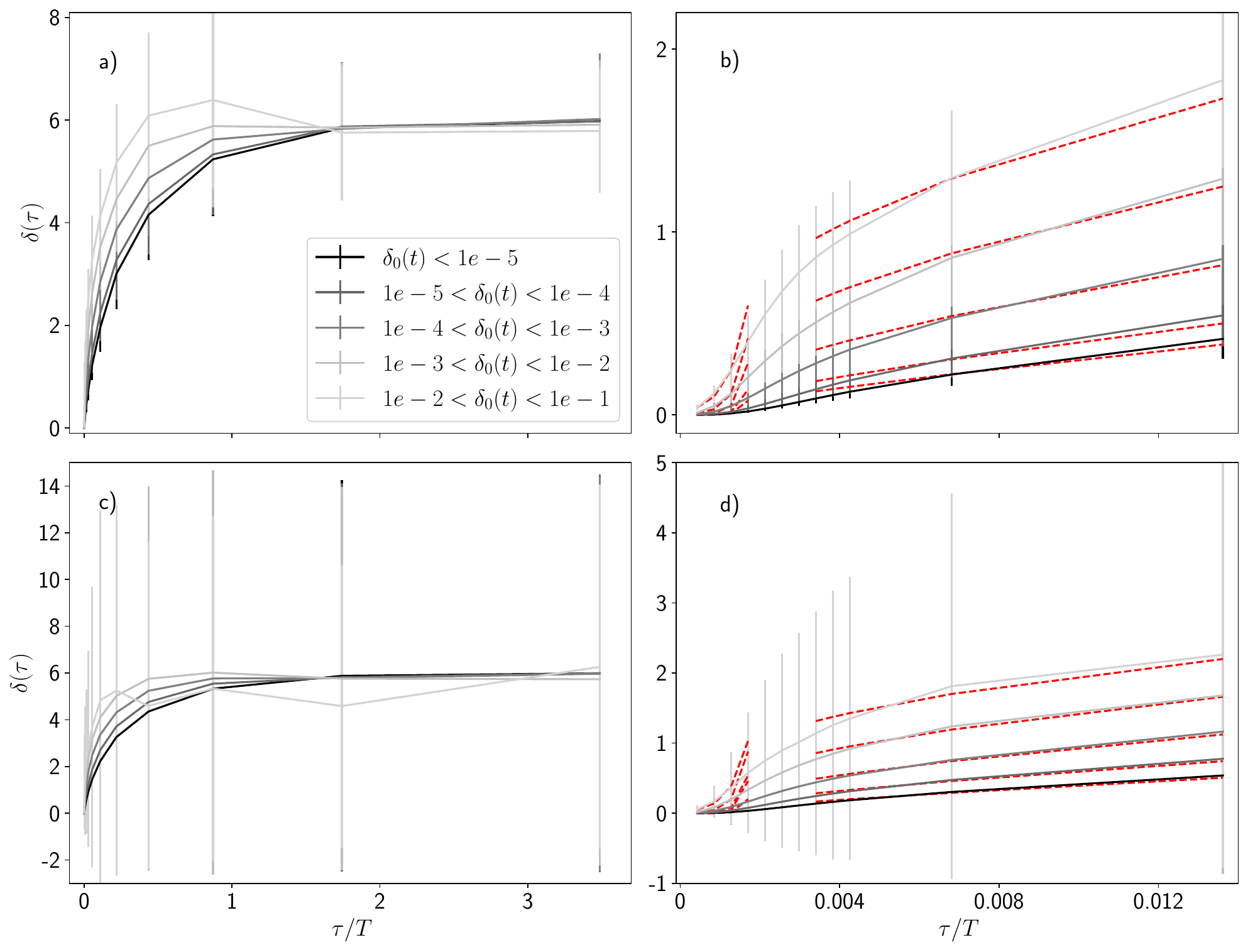}
\caption{ Error $\delta(\tau)$ as a function of $\tau/T$ for different values of the initial error $\delta_0$ and two values of $k$: Upper panel: \textbf{Case $\mathbf{k=50}$.} Lower panel: \textbf{Case $\mathbf{k=2}$.} Darker colors indicate smaller initial error $\delta_0$. In (a) and (c) for $\tau/T$ from $0$ to $3.5$. In (b) and (d) for very small $\tau/T \in [0,0.013]$. Red dashed lines in b) and (d) correspond to exponential and power law fits following equations (\ref{eq:Lyapunov}) and (\ref{eq:dissipation}) respectively. 
}
\label{fig:2}
\end{figure}

Figure~\ref{fig:2} a) and b) show the error $\delta(\tau)$ as a function of the time delay $\tau$ for different values of the initial error $\delta_0$ in the case $k=50$. In black for the smallest initial error $\delta_0<1e-5$ and in light gray for the largest initial error $1e-2<\delta_0<1e-1$. Figure~\ref{fig:2} a) shows a wide domain of $\tau/T \in [0,3.5]$ and Figure~\ref{fig:2} b) is a zoom at small $\tau/T \in [0,0.013]$ to highlight the different regimes of error growth. At long times, we observe the saturation stage, where the error becomes equal to $2v_{rms}^2$. Due to our normalization, the value $2v_{rms}^2$ corresponds to twice the dimension, i.e. $6$ here. This is the expected saturation value when analogues fully decorrelate. We observe that the mediocre analogues (large initial error) have a tendency to saturate at an earlier time than the excellent analogues (low initial error), in agreement with Lorenz's intuition. The error overshoot is just a statistical illusion, due to large error bars, as it decreases if we take instead a lower value of $k$, as in Figure~\ref{fig:2} c). Zooming into small values of $t$, we can observe the early stage of the error growth. We see two regimes in both cases $k=50$ and $k=2$: first, an exponential regime, which can be identified with the chaotic stage. Then, an algebraic stage follows, identified with the onset of spontaneous stochasticity. These observations are true for both values of $k$, but with higher error bars in the case $k=2$. They are also true for $p=10$ and $p=15$, but in these cases the chaotic regime extends to slightly larger times before the beginning of the algebraic one.

To confirm the interpretation of our observations and quantify the properties of analogues, we now explore in more detail the characteristics of the chaotic and algebraic stages.

\subsection{Extraction of characteristic quantities}

Following equations~(\ref{chaotic}) and~(\ref{SpSt}), we seek to fit the error growth at small times as:

\begin{eqnarray}
    \delta(\tau)  &=\delta_0 e^{\gamma \tau} \,\,\,\,\,&\text{for} \,\,\,\,\, \tau<\tau_c \label{eq:Lyapunov}  \\
     \delta(\tau) &= p\,C_{\alpha}\,u_{rms}^{2\left(1-\frac{3\alpha}{2}\right)}\,\epsilon^{\alpha}\,\bar{u}^{\alpha}\,(\tau+t_{\delta_0})^{\alpha}  \,\,\,\,\,&\text{for} \,\,\,\,\, \tau_c<\tau<\tau_s \label{eq:dissipation}
\end{eqnarray}

\noindent where $\gamma$ is the Lyapunov exponent, $\epsilon$ is the mean energy dissipation rate, $u_{rms}$ the rms velocity fluctuation, $\bar{u}$ the mean advection velocity, $C_{\alpha}$ a dimensionless constant, and $t_{\delta_0}$ depends upon when the chaotic stage ceases to be valid. The factor $p$ in Eq.~(\ref{eq:dissipation}) arises because $\delta$, as defined in Eq.~(\ref{eq:errordefsuc}), sums the contributions of the $p$ components of the embedding vector; since $\tau_*=dt\ll\tau_K$, these $p$ velocity increments are nearly identical, so that $\delta\simeq p\,\delta u^2$. It is the same factor that sets the saturation level at $2p\,v_{rms}^2$, and extracting it explicitly makes $C_{\alpha}$ a per-component amplitude, directly comparable to the usual structure-function constants.

The form of Eq.~(\ref{eq:dissipation}) is entirely fixed by two requirements. First, in our single-probe measurement the time delay $\tau$ is converted into a spatial separation only through Taylor's frozen-flow hypothesis, $r=\bar{u}\tau$; since a two-point statistic cannot depend on the frame of reference, $\bar{u}$ and $\tau$ may appear only through the product $\bar{u}\tau$. Second, dimensional analysis applied to the remaining parameters, $\epsilon$ and $u_{rms}$, then fixes the prefactor uniquely, leaving only the dimensionless constant $C_{\alpha}$ undetermined. Equivalently, in terms of the separation,

\begin{equation}
\delta(r)=p\,C_{\alpha}\,u_{rms}^{2\left(1-\frac{3\alpha}{2}\right)}\,(\epsilon r)^{\alpha},
\label{eq:dissipationspace}
\end{equation}

\noindent an expression in which the mean velocity has disappeared, as it must. For $\alpha=2/3$ the exponent of $u_{rms}$ vanishes and one recovers the Kolmogorov second-order structure function $\delta=C_{2/3}(\epsilon r)^{2/3}$; the deviation from Kolmogorov scaling thus appears as a pure power of the turbulence intensity, with an exponent that changes sign on crossing $\alpha=2/3$. We note in passing that writing the prefactor as $\epsilon^{\alpha}\bar{u}^{2(1-\alpha)}$, as dimensional analysis on $(\epsilon,\bar{u},\tau)$ alone would suggest, leaves a residual $\bar{u}^{2-3\alpha}$ in Eq.~(\ref{eq:dissipationspace}) and is therefore admissible only at $\alpha=2/3$.

Considering $\alpha=2h$, $\alpha=2/3$ is then consistent with the usual H\"older exponent $h=1/3$. The power law in this regime is different from the value $\alpha=\frac{2h}{1-h}$, obtained in section~\ref{sec:spont}, due to the mean flow advection in our dataset. In the normalized variables used throughout this paper, in which $\delta$ is measured in units of $u_{rms}^2$ and $\tau$ in units of the integral time $T$, Eq.~(\ref{eq:dissipation}) reduces to $\delta=p\,C_{\alpha}\left[(\tau+t_{\delta_0})/T\right]^{\alpha}$, so that $C_{\alpha}$ and $\alpha$ are read directly from the data of Fig.~\ref{fig:2} without any further calibration.

\begin{figure}[ht!]
\includegraphics[width=\linewidth]{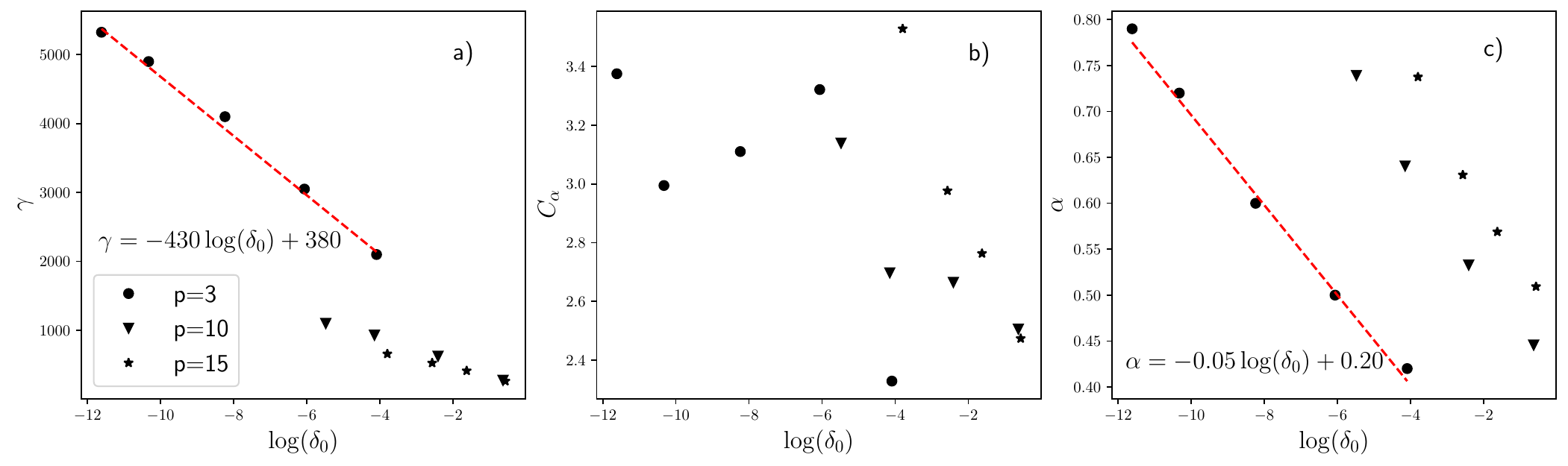}
 \centering
\caption{a) Lyapunov exponent $\gamma$, b) dimensionless constant $C_{\alpha}$ of Eq.~(\ref{eq:dissipation}) and c) exponent $\alpha$, all as a function of the logarithm of the initial error $\log(\delta_0)$ and for reconstructed phase spaces of dimension $p=3$, $p=10$ and $p=15$. The red dashed lines are linear fits of the measures corresponding to $p=3$; no fit is shown in panel b), where $C_{\alpha}$ displays no significant trend.}\label{fig:3}
\end{figure}

Figure~\ref{fig:3} shows the values of the Lyapunov exponent $\gamma$, the dimensionless constant $C_{\alpha}$ and the exponent $\alpha$ estimated by fitting the behavior of $\delta(\tau)$ as a function of $\tau/T$ with equations~(\ref{eq:Lyapunov}) for the Lyapunov exponent and~(\ref{eq:dissipation}) for $C_{\alpha}$ and $\alpha$, and for different initial errors $\delta_0$. Estimations have been performed in reconstructed phase spaces of dimensions $p=3$ (circles), $p=10$ (triangles) and $p=15$ (stars).

Quantitatively, $\alpha$ decreases from $\alpha \sim 0.79$ for the best analogues to $\alpha \sim 0.42$ for the most mediocre ones, the Kolmogorov value $\alpha = 2/3$ being recovered around $\delta_0 \sim 1e-4$. Considering $\alpha = 2h$, this corresponds to an effective H\"older exponent decreasing from $h \sim 0.39$ to $h \sim 0.20$, a range that coincides with the usual multifractal support of the velocity field and the Kolmogorov value $h = 1/3$. The quality of an analogue therefore somehow acts as a filter on the local regularity of the signal: a small $\delta_0$ can only be reached for configurations that recur frequently in the record, that is for typical, moderately fluctuating segments, whereas the large-$\delta_0$ bins are dominated by states with few close neighbours (the rare, intense events, which are also the most singular ones). The measured $\alpha(\delta_0)$ is thus a direct readout of that selection and consistent with the intermittency bias discussed above.

Independently of the dimension, the Lyapunov exponent and the exponent $\alpha$ are decreasing functions of $\delta_0$, with approximate logarithmic scaling in the initial error $\delta_0$, whereas the dimensionless constant $C_{\alpha}$ shows no systematic trend and remains of the same order over the whole range, $C_{\alpha}\sim 2.4-3.4$. This last point is expected: $C_{\alpha}$ characterizes the flow rather than the way it is sampled, and it should therefore be insensitive to the quality of the analogue. Its magnitude is also the expected one: in the bin where $\alpha=2/3$, Eq.~(\ref{eq:dissipationspace}) reduces to $p$ times the second-order structure function of the velocity field, so that $C_{\alpha}$ should there coincide with the Kolmogorov constant for longitudinal structure functions, $C_2\simeq 2$. The measured value, $C_{2/3}\simeq 3.0$, is somewhat larger but of the right order, which supports the identification of the algebraic stage with the Kolmogorov scaling of the velocity increments.
In an ideal case where the density of the phase space is homogeneous and the analogues can be sampled in an equivalent way independently of their position in the phase space, the exponent $\alpha$ and the Lyapunov exponent should not depend on $\delta_0$ and should correspond to the Hurst exponent and the typical Lyapunov exponent of the fluid. This is the case for Holder continuous processes with a single Holder exponent, single value of $\alpha$, such as fractional Brownian motion~\cite{GraneroBelinchon2026, GraneroBelinchon2026a}. We can then interpret this dependence as a signature of the bias induced by the inhomogeneity of the reconstructed phase space due to intermittency, which impacts the selection of analogues for a given initial error. Indeed, rare events will typically have fewer analogues than typical events, and will only correspond to situations with a large initial error. Therefore, the higher the initial error, the higher the weight of rare events. If this interpretation is correct, it means that rare events are characterized by a smaller $\alpha$ and a smaller Lyapunov exponent than typical events. It would be interesting to confirm such results with appropriate numerical simulations. The dependence of $\lambda$ on $\delta_0$ is nonetheless consistent with the theoretical picture of Crisanti et al.~\cite{crisantiIntermittencyPredictabilityTurbulence1993, crisantiPredictabilityVelocityTemperature1993}, who showed, within the multifractal formalism, that the Lyapunov exponent of an intermittent flow is not a single value but a statistically distributed quantity, whose variance grows with the Reynolds number precisely because of the contribution of rare events; the smaller exponents we observe for larger $\delta_0$ can then be read as an experimental signature of the low-exponent tail of that distribution.

\section{Conclusion}
We have demonstrated that Lorenz's concept of analogues can be effectively applied to turbulent flows to investigate error growth and extract information about chaoticity and spontaneous stochasticity from turbulent signals. In qualitative agreement with Lorenz's original findings, we observe that analogues with larger initial errors saturate more rapidly than those with smaller initial errors, validating his intuition.

On a quantitative level, we extracted the equivalent of a Lyapunov exponent in the early chaotic stage and, in the subsequent algebraic stage, the power law exponent $\alpha$ together with the dimensionless amplitude $C_{\alpha}$ of Eq.~(\ref{eq:dissipation}). However, the dependence of the Lyapunov exponent and of the power law exponent of the spontaneous stochasticity regime on the initial error suggests the presence of a statistical bias linked to the intermittency of turbulence. This bias arises because rare events, which are less densely represented in phase space, are preferentially sampled when considering larger initial errors. Our results indicate that these rare events are characterized by a smaller Lyapunov exponent and a smaller $\alpha$ exponent compared to typical events, while the amplitude $C_{\alpha}$, which characterizes the flow rather than its sampling, remains of the same order throughout.

While the decrease of $\alpha$ (that indicates the selection of increasingly singular events as $\delta_0$ grows) is the expected signature of intermittency, the concomitant decrease of the Lyapunov exponent is less intuitive and warrants further investigation. Future work could explore this bias in greater detail and attempt to connect it with other theories of rare events, such as multifractal theory. Additionally, while our study is currently limited to $0+1$-dimensional data, it would be valuable to extend it to scalar $2+1$-dimensional datasets (such as pressure maps in climate or fluid experiments), where analogues can still be identified without encountering the curse of dimensionality~\cite{Platzer2021, Platzer2021a}.

\begin{acknowledgments}
 This work received funding from ANR BANG grant agreement no. ANR-22-CE30-0025. This work was also supported by a French government grant managed by the Agence Nationale de la Recherche France 2030 under the reference ANR-24-RRII-0001.     
    \end{acknowledgments}   
    
\bibliographystyle{unsrt}
\bibliography{biblio.bib}

\end{document}